\documentclass[%
 aip,
 floatfix,amsmath,amssymb,
 reprint,%
nofootinbib]{revtex4-1}

\usepackage{graphicx}
\usepackage{dcolumn}
\usepackage{bm}
\usepackage[utf8]{inputenc}
\usepackage[T1]{fontenc}
\usepackage{mathptmx}
\DeclareMathAlphabet{\mathcal}{OMS}{cmsy}{m}{n}
\usepackage{etoolbox}
\usepackage{bm}
\usepackage{ulem}

\usepackage{siunitx}
\usepackage{amsmath}
\usepackage{braket}
\usepackage[hang,flushmargin,symbol]{footmisc}
\usepackage{xcolor}
\usepackage{placeins}
\usepackage{longtable}
\usepackage{natbib}
\newcommand{\mainMgFTrans}{A$^2\Pi \leftarrow $X$^2\Sigma^+$ }

\newcommand{\mainMgFTransUpper}{A$^2\Pi_{3/2} \leftarrow $X$^2\Sigma^+$ }
\newcommand{\mainMgFTransLower}{A$^2\Pi_{1/2} \leftarrow $X$^2\Sigma^+$ }
\newcommand{\MgFgdstate}{X$^2\Sigma^+$ }
\newcommand{\MgFexstateUpper}{A$^2\Pi_{3/2}$ }
\newcommand{\MgFexstateLower}{A$^2\Pi_{1/2}$ }
\newcommand{\MgFexstate}{A$^2\Pi$ }
\newcommand{\expect}[1]{\langle #1 \rangle}
\newcommand{\MgFisoA}{^{24}\mathrm{MgF}}
\newcommand{\MgFisoB}{^{26}\mathrm{MgF}}

\makeatletter
\def\@email#1#2{%
 \endgroup
 \patchcmd{\titleblock@produce}
  {\frontmatter@RRAPformat}
  {\frontmatter@RRAPformat{\produce@RRAP{*#1\href{mailto:#2}{#2}}}\frontmatter@RRAPformat}
  {}{}
}%
\makeatother
\begin{document}

\preprint{AIP/123-QED}

\title[Hyperfine resolved optical spectroscopy of the \mainMgFTrans transition in MgF]{Hyperfine resolved optical spectroscopy of the \mainMgFTrans transition in MgF}
\author{M. Doppelbauer}
\altaffiliation{These authors contributed equally to this work}
\author{S. C. Wright}
\altaffiliation{These authors contributed equally to this work}
\author{S. Hofsäss}
\affiliation{Fritz-Haber-Institut der Max-Planck-Gesellschaft, Faradayweg 4-6, 14195 Berlin, Germany}
\author{B. G. Sartakov}
\affiliation{General Physics Institute, Russian Academy of Sciences, Vavilovstreet 38, 119991 Moscow, Russia}
\author{G. Meijer}
\email{meijer@fhi-berlin.mpg.de}
\author{S. Truppe}
\email{truppe@fhi-berlin.mpg.de}
\affiliation{Fritz-Haber-Institut der Max-Planck-Gesellschaft, Faradayweg 4-6, 14195 Berlin, Germany}

\date{\today}

\begin{abstract}
We report on hyperfine-resolved laser spectroscopy of the \mainMgFTrans transition of MgF, relevant for laser cooling. We recorded 25 rotational transitions with an absolute accuracy of better than \SI{20}{\mega\hertz}, assigned 56 hyperfine lines and determined precise rotational, fine and hyperfine structure parameters for the \MgFexstate state. The radiative lifetime of the \MgFexstate state was determined to be 7.2(3) ns, in good agreement with \textit{ab initio} calculations. The transition isotope shift between bosonic isotopologues of the molecule is recorded and compared to predicted values within the Born-Oppenheimer approximation. We measured the Stark effect of selected rotational lines of the \mainMgFTrans transition by applying electric fields of up to \SI{10.6}{\kilo\volt\per\centi\meter} and determined the permanent electric dipole moments of $^{24}$MgF in its ground \MgFgdstate and first excited \MgFexstate states to be $\mu_X=\SI{2.88(20)}{D}$ and $\mu_A=\SI{3.20(22)}{D}$, respectively. Based on these measurements, we caution for potential losses from the optical cycling transition, due to electric field induced parity mixing in the excited state. In order to scatter $10^4$ photons, the electric field must be controlled to below \SI{1}{\volt\per\centi\meter}. 
\end{abstract}

\maketitle

\section{Introduction}
Recently, magnesium monofluoride (MgF) has been identified as a promising candidate molecule for laser cooling and magneto-optical trapping experiments \cite{Kang2015, Xu2016, Chae2021}. Compared to other group II monofluoride molecules that have been laser-cooled so far\cite{Barry2014,Truppe2017,Anderegg2017a}, MgF is lighter and has a stronger optical cycling transition in the ultraviolet. These properties allow for exerting a large radiation force to rapidly slow and cool the molecules and produce a magneto-optical trap with a high capture velocity. The predicted low off-diagonal vibrational branching and the simple hyperfine structure of MgF reduces the complexity of the optical setup significantly. 

Rotationally resolved optical spectra of MgF have been recorded in absorption \cite{Jenkins1934, Fowler1941, Barrow1967, Walker1968,Novikov1971} and emission \cite{Novikov1971}. The vibration-rotation emission in the electronic ground state has also been studied \cite{Barber1995}. Precise hyperfine and rotational constants of MgF in its \MgFgdstate electronic ground state for vibrational states $v=0,1,2$ and 3 were determined from its millimeter-wave spectrum \cite{Anderson1994,Anderson1994a}. Recently, Xu et al. \cite{Xu2019} recorded optical absorption spectra of the \mainMgFTrans transition, resolving a prior debate in the literature about the sign and value of the spin-orbit coupling constant. However, this study suffered from a large systematic frequency offset of about \SI{-4.1}{\giga\hertz} and the Hamiltonian that was used to fit to the experimental data did not account for the presence of $\Lambda$-doubling and hyperfine structure (\textit{vide infra}) in the excited state\footnote[2]{A further study from the same group \cite{Gu2021} was submitted during the preparation of this article. We find a systematic frequency offset of about $+$\SI{2.3}{GHz} in the reported line centers.}. Recently, optical cycling experiments have been performed \cite{Xia2021}, a first step towards laser cooling experiments.

MgF has also been studied theoretically using ab initio methods \cite{Walker1970, Kang2015, Xu2016}. Pelegrini \textit{et al.} \cite{Pelegrini2005} calculated various properties of MgF as part of a wider study of group II monofluorides. These predictions show good agreement with the available experimental data for CaF. For MgF, they predicted a radiative lifetime of \SI{7.16}{\nano\second} for the \MgFexstate,$v'=0$ level, and a decay probability of 1.4\% to the \MgFgdstate, $v''=1$ first vibrationally excited state. They also predicted the permanent electric dipole moments for the ground and excited states to be \SI{2.67}{D} and \SI{4.23}{D} respectively.

Here, we present hyperfine-resolved UV laser-induced fluorescence (LIF) spectra of MgF produced in a cryogenic buffer gas molecular beam. The large frequency calibration error present in the previous study by Xu \textit{et al.}\cite{Xu2019} is corrected by calibrating our wavemeter with known transition frequencies in Yb. The eigenvalues of an appropriate Hamiltonian are fitted to the measured hyperfine energy levels to derive precise spectroscopic constants for the \MgFexstate state. We record and analyse transition isotope shifts between the two bosonic isotopologues and compare to predictions from mass scaling arguments. The spectral width of isolated lines is measured with high accuracy to determine the radiative lifetime of the \MgFexstate state. The electric dipole moments of the ground and excited states are deduced from the Stark shifting of individual rotational lines in electric fields of up to \SI{10.6}{\kilo\volt\per\centi\meter}. We then determine how opposite parity levels in the excited state mix in an external electric field. This effect can result in large losses from the optical cycling scheme, if stray electric fields are not well-controlled.

The group II metal monofluorides are interesting candidates for laser cooling because of the single, unpaired, and metal-centered electron that is polarized away from the fluorine atom. The internuclear distance and potential energy curves of the ground and first excited states are very similar. This leads to a very diagonal Franck-Condon matrix which reduces the number of vibrational repump lasers required to scatter a large number of photons. Measurements of the fine and hyperfine structure and the electric dipole moments provide information about the spin density at the fluorine nucleus and the charge distribution. This provides information to better understand the bonding structure in these molecules \cite{Kaendler1989}. We compare our results to the other group II monofluoride molecules, CaF, SrF and BaF, which have been studied in detail.

\section{Hamiltonian}
We use the following effective Hamiltonian, which operates in a given vibrational state with energy $E_0$:
\begin{equation}\label{eq:hamil}
    \begin{aligned}[b]
        H&=B\mathbf{N^2}-D\mathbf{N^4}+A \mathbf{L}\cdot \mathbf{S}+\gamma\mathbf{N}\cdot\mathbf{S}\\
        &-\frac{1}{2}p(N_+S_++N_-S_-)+\frac{1}{2}q(N_+^2+N_-^2)\\
        &+a L_zI_z+b_F\mathbf{S}\cdot\mathbf{I}+\frac{1}{3}c(3S_zI_z-\mathbf{S}\cdot\mathbf{I})\\
        &-\frac{1}{2}d(S_+I_++S_-I_-)
    \end{aligned}
\end{equation}
 It describes rotation ($B$, $D$), spin-orbit $A$, spin-rotation $\gamma$, $\Lambda$-doubling ($p$, $q$) and magnetic hyperfine interactions ($a$, $b_F$, $c$ and $d$). The ground state of MgF is a \MgFgdstate state, for which $\Lambda=0$ and therefore $A=p=q=a=d=0$.
 
\begin{figure}
    \centering
    \includegraphics[]{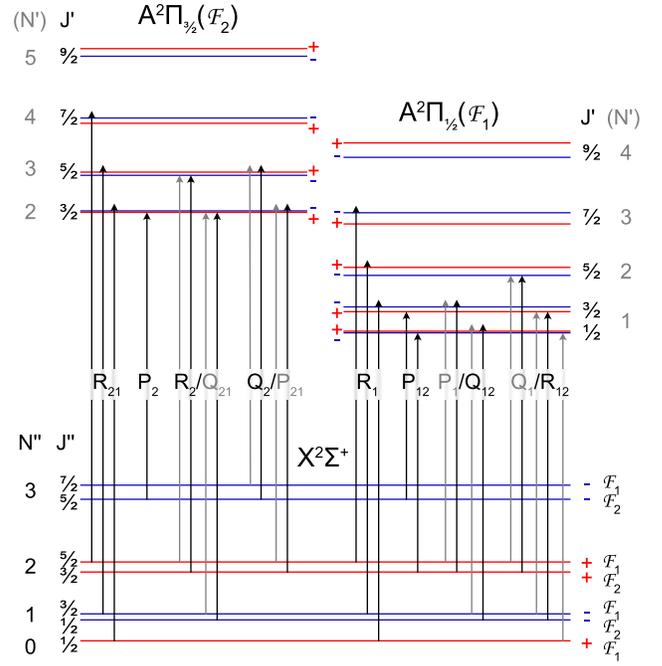}
    \caption{Schematic energy level diagram for a regular $^2\Pi -^2\Sigma^+$ system, where $p$ and $q$ are both positive. Positive parity states are shown in red while negative parity states are shown blue (not to scale). $N'$ is not well defined for low-$J'$. For clarity, where the spacings are small, arrowheads extending above (below) a doublet pair mark a transition to the upper (lower) doublet.}
    \label{fig:biglevelscheme}
\end{figure}

 Fig. \ref{fig:biglevelscheme} shows the relevant levels and transitions in absence of hyperfine structure. For a given value of $J$, the level with the lowest energy is labeled $\mathcal{F}_1$. We use $\Delta J_{\mathcal{F'}\mathcal{F}''}(N'')$ to label the transitions. The energies of the $J$-levels in the ground state can be calculated using,
 \begin{equation}
    \begin{aligned}[b]
    E(N)&=E_0+BN(N+1)+\gamma N/2 &  \textrm{for}    &&  J&=N+1/2\\
    E(N)&=E_0+BN(N+1)-\gamma (N+1)/2 & \textrm{for} &&  J&=N-1/2 ,
    \end{aligned}
    \label{eq:XstateEnergies}
\end{equation}
and in the A$^2\Pi$ state using the following formula,
 \begin{equation}
    \begin{aligned}[b]
        E(J)&=E_0-\frac{1}{2}\gamma+B\left(J(J+1)
        -\frac{3}{4}\right)\\
        &\mp\sqrt{\left(\frac{1}{2}(A+\gamma)-B\right)^2+
    \left(B-\frac{1}{2}\gamma\right)^2\left(J(J+1)-\frac{3}{4}\right)} \hspace{0.2cm}.
    \end{aligned}
    \label{eq:AstateEnergies}
\end{equation}
Here, the minus (plus) sign applies for the $\mathcal{F}_1(\mathcal{F}_2)$ levels respectively.
In the case of $J'=1/2$, expression \eqref{eq:AstateEnergies} reduces to
\begin{equation}
    E\left(J'=\frac{1}{2}\right)=E_0-\frac{1}{2}\gamma-\frac{1}{2}(A+\gamma)+
    B.
\end{equation}


\noindent When comparing transition frequencies of different isotopologues, an explicit form for the energy $E_0$ is required. Here we assume that

\begin{equation}
    E_0 = T_e + \omega_e(v+1/2) - \omega_ex_e(v+1/2)^2,
    \label{eqn:E0definition}
\end{equation}

\noindent with $T_e$ being the potential energy minimum of an electronic state, and the remaining terms describing the vibrational energy up to second order in the vibrational quantum number $v$. 

\section{Experimental setup}
\label{experimental}

\begin{figure}
    \centering
    \includegraphics[]{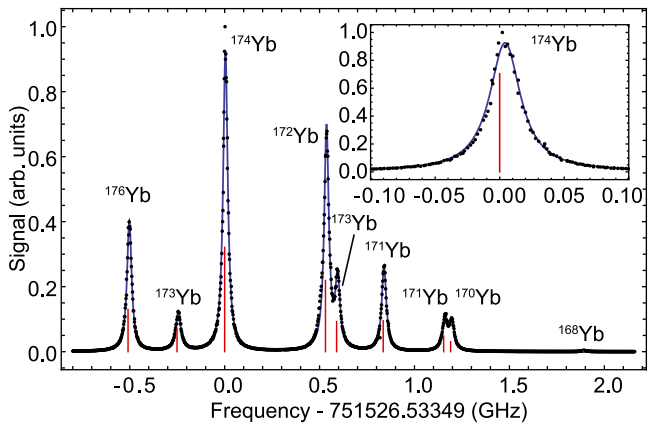}
    \caption{Laser induced fluorescence spectrum of the $(6s6p)^1P_1\leftarrow(6s^2)^1S_0$ transition in Yb used to verify the accuracy of our wavemeter. The black dots represent the data recorded in this study and the blue curve shows a fit using Lorentzian lineshapes. The red sticks represent the line positions obtained by Kleinert \textit{et al.}\cite{Kleinert2016} with an absolute accuracy of better than \SI{1}{\mega\hertz}. The inset shows the spectral line from the $^{174}$Yb isotope in more detail.}
    \label{fig:Ybspectrum}
\end{figure}

The spectrometer used for this study is similar to the one described previously \cite{Doppelbauer2021,Hofsass2021}. MgF molecules are produced in a cryogenic helium buffer gas cell that is cooled to \SI{2.7}{K} using a closed-cycle helium cryocooler. The cell's geometry is based on the design of Truppe \textit{et al.} \cite{Truppe2018}; it has a length of \SI{40}{mm} with a bore diameter of \SI{10}{mm}, and an aperture of \SI{4}{mm}. A Mg rod is ablated by \SI{20}{mJ} of pulsed Nd:YAG laser (Continuum Minilite II, 1064 nm) radiation focused to a waist diameter of 0.4 mm. The hot Mg atoms react with NF$_3$ gas (\SI{0.001}{sccm} flow rate, 100 K) to form MgF molecules. The molecules are cooled by collisions with the cryogenic He buffer gas which is flowing into the cell continuously at a rate of \SI{1}{sccm}. This helium flow also extracts the pulse of molecules from the cell into a molecular beam with a rotational temperature of about 4 K and a mean forward velocity of typically \SI{160}{\meter\per\second}. The forward velocity increases over several thousand ablation shots, but can be restored to its original value by cleaning the cell. The molecules are detected by laser-induced fluorescence (LIF) \SI{44}{cm} downstream from the buffer gas cell aperture \cite{Doppelbauer2021,Hofsass2021}. The transverse velocity spread of the molecular beam is reduced to about \SI{1}{\meter\per\second} by a 2x\SI{2}{\square\milli\meter} square aperture placed at the entrance to the LIF detector. Here, a continuous wave (CW) \SI{359}{nm} laser beam from the second harmonic of a titanium sapphire laser intersects the molecular beam perpendicularly. We ensure that any Doppler shift arising from misalignment of the probe beam is below \SI{10}{MHz} by measuring spectra with and without retroreflecting the laser. The LIF is imaged onto a photomultiplier tube (PMT, Hamamatsu R928) and the resulting photo-current is amplified to give a time-dependent fluorescence signal. We measure and stabilize the fundamental wavelength of the titanium sapphire laser using a wavemeter (HighFinesse WS8-10), which has an absolute accuracy of \SI{20}{\mega\hertz} and a measurement resolution of \SI{0.4}{\mega\hertz}. The wavemeter is calibrated using a temperature-stabilized HeNe laser (SIOS), whose absolute frequency is known to within \SI{5}{\mega\hertz}. Additionally, we determine the $(6s6p)^1P_1\leftarrow(6s^2)^1S_0$ transition frequency and isotope shifts of Yb by applying high-resolution laser spectroscopy to a pulsed buffer gas beam of Yb atoms. These frequencies are known with an absolute and relative uncertainty of better than \SI{1}{\mega\hertz} \cite{Kleinert2016}. Our experimental spectrum is presented in Fig. \ref{fig:Ybspectrum}. The lines under the spectrum show the transition frequencies measured by Kleinert \textit{et al.}\cite{Kleinert2016}. In our spectrum, the line-centers are determined from a fit to multiple Lorentzian lineshapes and listed in Table \ref{tab:Ybtable}. We reproduce the absolute transition frequencies within \SI{10}{MHz} and the relative frequencies (isotope shifts) to within \SI{1}{MHz} over the range of several GHz. Since the deviation between our measured line centers and the published values is within the absolute accuracy of the wavemeter, we use the calibration from the HeNe laser without further correction. We found it was necessary to add a constant flow of dry nitrogen gas through the doubling cavity, in order to avoid absorption of the fundamental light by water vapor. This effect was particularly pronounced near the Q$_1$(0) fundamental
frequency of \SI{834.3255}{\tera\hertz}. 

\begin{table}
\caption{\label{tab:Ybtable} The measured $(6s6p)^1P_1\leftarrow(6s^2)^1S_0$ transition frequencies of Yb relative to the transition frequency of $^{174}$Yb as determined by Kleinert \textit{et al.} \cite{Kleinert2016}. For the isotopes with a nuclear spin of $I\neq0$, $F'$ is given in brackets. The last column gives the absolute frequency differences of the isotope shifts between the two measurements with a standard deviation (SD) of \SI{1.3}{\mega\hertz}.}
\begin{ruledtabular}
\begin{tabular}{lccc}
Isotope&Isotope shift (MHz)\cite{Kleinert2016}&This study (MHz)&$\Delta f$\\
\hline
176 & -508.89$\pm$0.09 & -502.11$\pm$0.13 & 6.78\\
173 ($F'=5/2$) & -250.78$\pm$0.33 & -243.79$\pm$0.43 & 6.99\\
174 & 0 & 4.16$\pm$0.05 & 4.16 \\
172 & 531.11$\pm$0.09 & 536.51$\pm$0.08 & 5.40\\
173 ($F'=7/2$) & 589.75$\pm$0.24 & 595.14$\pm$0.24 & 5.39\\
171 ($F'=3/2$) & 835.19$\pm$0.20 & 839.57$\pm$0.20 & 4.38\\
171 ($F'=1/2$) & 1153.68$\pm$0.25 & 1160.96$\pm$0.54 & 7.28\\
170 & 1190.36$\pm$0.49 & 1196.78$\pm$0.63 & 6.42\\
168 & 1888.80$\pm$0.11 & 1892.44$\pm$0.11 & 3.64\\
\end{tabular}
\end{ruledtabular}
\end{table}

To determine the electric dipole moment of MgF in the \MgFexstate and \MgFgdstate states, we install transparent copper mesh electrodes below and above the molecular beam to apply electric fields to the molecules inside the LIF detector. The distance between the electrodes is measured to be \SI{9.0(3)}{\milli\meter}. The voltage on the electrodes is supplied by a high-voltage power supply (Spellman SL1200) and measured with a calibrated high-voltage probe and multimeter with a combined relative accuracy of 10$^{-4}$.


\section{Isotope shifts, spectroscopic constants, hyperfine structure, and $\Lambda$-doubling}
The vibrational, rotational, fine and hyperfine structure of the ground electronic state of MgF is well known. To improve the spectroscopic parameters for the rotational, fine and hyperfine structure of the \MgFexstate state, we record low-$J$ rotational lines of the \mainMgFTrans transition. The isotope shifts between the $^{26}$MgF and $^{24}$MgF are discussed in section \ref{sec:isotopeShifts}. From section \ref{sec:hyperfineStructure} onwards, we focus on the most abundant $^{24}$MgF, summarizing the results of our measurements, and point out some important differences with the other group II monofluorides.

\subsection{Isotope shifts}
\label{sec:isotopeShifts}

We use a Mg metal ablation target with a natural isotopic abundance of 79\%, 10\% and 11\% for $^{24}$Mg, $^{25}$Mg, and $^{26}$Mg, respectively. $^{24}$Mg and $^{26}$Mg are bosons with a nuclear spin $I(^{24}\textrm{Mg})=0$ and $I(^{26}\textrm{Mg})=0$ whereas $^{25}$Mg is a fermion with a nuclear spin $I(^{25}\textrm{Mg})=5/2$. Fluorine has one stable isotope with a nuclear spin of $I(^{19}\textrm{F})=1/2$. 

\begin{figure}
    \centering
    \includegraphics{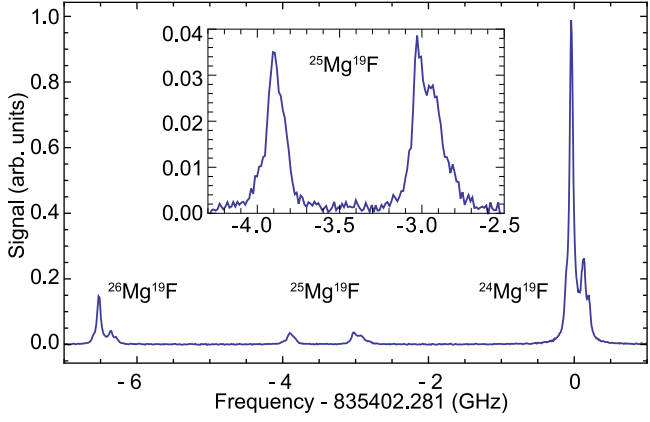}
    \caption{Isotope shift of the R$_2$/Q$_{21}$(1) line of MgF. The $^{26}$Mg and $^{24}$Mg isotopes are bosons with $I=0$, while $^{25}$Mg is a fermion with $I=5/2$ and has a more complex hyperfine structure. A zoom-in on the $^{25}$MgF isotope is shown as an inset.}
    \label{fig:R22isotopes}
\end{figure}

Fig. \ref{fig:R22isotopes} shows a typical spectrum when exciting the R$_2$/Q$_{21}$(1) line. In this example we scan over the three MgF isotopologues, and observe isotope shifts of \SI{-3.35}{GHz} ($^{25}$MgF) and \SI{-6.48}{GHz} ($^{26}$MgF), relative to $^{24}$MgF. The inset of Fig. \ref{fig:R22isotopes} shows the more complex hyperfine structure of $^{25}$MgF. Each rotational line is split into two groups of hyperfine lines separated by about 1GHz, which are both further split by a few hundred MHz. The larger splitting arises from the Fermi interaction between the electron and $^{25}$Mg nuclear spin in the ground electronic state\cite{Anderson1994}. We did not analyse the excited state in detail in this study.

The shift in transition frequencies between isotopologues is characteristic to a molecular species. This molecular isotope shift can be used as an additional means of identification, and to reveal small deviations from the Born-Oppenheimer approximation. Since $^{24}$MgF and $^{26}$MgF exhibit the same hyperfine structure, the shift in the gravity center of a rotational line can be found straightforwardly by comparing the positions of equivalent hyperfine peaks. We define the transition isotope shift, $\delta\nu_{i}$, as,

\begin{equation}
\delta\nu_{i} = \nu_i(^{26}\mathrm{MgF}) - \nu_i(^{24}\mathrm{MgF})
\end{equation}

\noindent with $\nu_i(j)$ being the absolute frequency of the optical transition $i$ in isotopologue $j$. Within the Born-Oppenheimer approximation, the isotope shift comes about through changes to the relevant reduced masses in the molecular system. For the rovibrational constants, the relevant reduced mass is that computed from the two atomic masses, $m_{\mathrm{Mg}}$ and $m_{\mathrm{F}}$, 
\begin{equation}
m_{mol} = \frac{m_{\mathrm{Mg}}m_{\mathrm{F}}}{m_{\mathrm{Mg}}+m_{\mathrm{F}}}\hspace{0.3cm}.
\end{equation}
For the electronic contribution $T_e$, it is that of the valence electron mass, $m_e$, and the remaining molecular mass,

\begin{equation}
 m_{el} = \frac{m_e(m_{\mathrm{MgF}}-m_e)}{m_{\mathrm{MgF}}} \hspace{0.3cm}.
\end{equation}

\noindent We define $\rho = \sqrt{m_{mol}(\MgFisoA)/m_{mol}(\MgFisoB)}$, which has the value of about $0.983$, and $\rho_{el} = m_{el}(^{24}\mathrm{MgF})/m_{el}(^{26}\mathrm{MgF})$, noting that $1-\rho_{el} = 5.67\times 10^{-7}$. To predict the isotope shifts, we use equations \eqref{eq:XstateEnergies}, \eqref{eq:AstateEnergies} and \eqref{eqn:E0definition} to calculate the energy differences, applying the following relations, 
\begin{equation}
\begin{gathered}
B^* = \rho^2 B,  \hspace{0.1cm}\\
\omega_e^* = \rho \omega_e, \hspace{0.1cm}
\omega_ex_e^* = \rho^2 \omega_ex_e, \\
T_e^* = \rho_{el}^{-1}T_e, \hspace{0.2cm}  
\label{eqn:isotopeShiftRelations}
\end{gathered}
\end{equation}

\noindent where the asterisks refer to the constants for $\MgFisoB$. The constants $\omega_e, \omega_ex_e$ are taken from the work of Novikov and Gurvich \cite{Novikov1971}\footnote[2]{While Barber \textit{et al.}\cite{Barber1995} provided more accurate values for the ground state, it is only the difference in ground and excited state constants that matters.}. The remaining values are taken from tables \ref{tab:gdstateparams} and \ref{tab:exstateparams} in the subsequent sections of this article. From the difference in vibrational constants we expect an isotope shift of \SI{-7.33}{GHz}, while the change in $T_e$ contributes $+$\SI{0.47}{GHz}.  

\begin{figure}
	\includegraphics[]{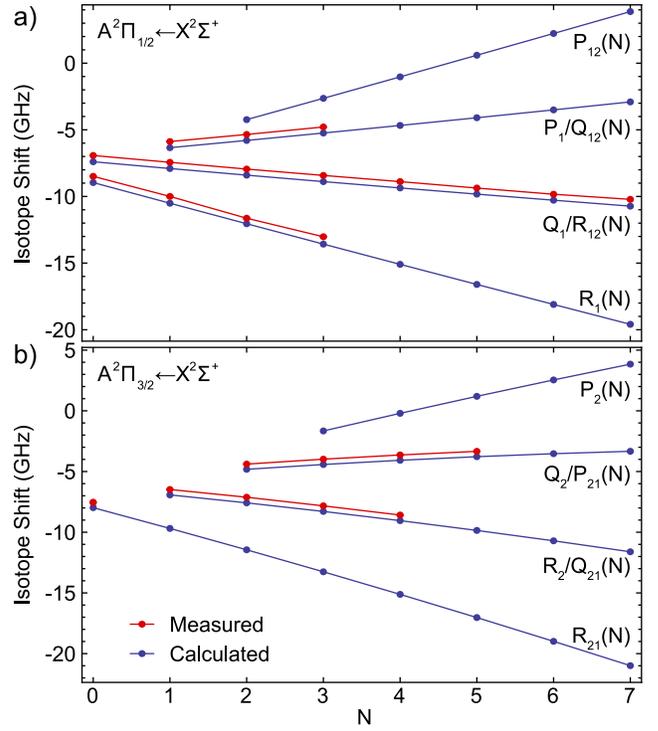} 
	\caption{Isotope shifts of the \mainMgFTrans transition in MgF. The blue joined points are calculated as described in the text, and the red joined points are from our dataset. (a) The \mainMgFTransLower branches. (b) The \mainMgFTransUpper branches.} 
	\label{fig:isotopeShifts}
\end{figure}

In Fig. \ref{fig:isotopeShifts} a and b, we plot the calculated $\delta\nu_i$ values for the four branches of the \mainMgFTransLower and \mainMgFTransUpper transitions. We plot the predicted values up to $N=7$, and compare with those available from our measurements. We observe a systematic difference between the measured and calculated values of $450-500$~MHz; the mean of the differences is 470~MHz and their standard deviation is 30~MHz. Thus, while the rotational dependence of the isotope shift is well described by relations \eqref{eqn:isotopeShiftRelations}, the combined shift of the electronic and vibrational terms is not. An inaccuracy of 1.8~cm$^{-1}$ in any one of the vibrational constants would be required to account for our observations, which seems unlikely. According to Hougen\cite{Hougen1970}, we should also include a term $B\expect{L_{\perp}^2}$ in the definition of $E_0$, which accounts for the component of the electronic orbital angular momentum perpendicular to the internuclear axis. A non-zero value of $\expect{L_{\perp}^2}$ would increase the discrepancy, by up to 540~MHz. It is therefore likely that the specific mass shift contribution to $T_e$, the field shift of the Mg nucleus, or deviations from the Born-Oppenheimer approximation are responsible for the additional shift. These can only be derived by more sophisticated calculations, and our values provide an important benchmark in this regard.

\subsection{Spectroscopic constants of the \MgFexstate state}
\label{sec:hyperfineStructure}
To determine the spin-orbit, rotational, spin-rotation, $\Lambda$-doubling, and hyperfine constants of the \MgFexstate state, we record 25 hyperfine-resolved rotational lines of the \mainMgFTransLower and \mainMgFTransUpper transitions. The lines are slightly broadened by a small, uncompensated ambient magnetic field of 0.8 G in the LIF detector, and the effect of optical pumping between hyperfine and rotational states (discussed in section \ref{sec:lifetime}); they are fitted using a sum of Lorentzian lineshapes. Using a Voigt profile did not change the fit residuals significantly. In this way, we determine the centers of 56 hyperfine lines for $N''\leq 4$, the observed frequencies of which are listed in Table \ref{tab:linelist} of the Appendix. These line centers were used in a least-squares fit to determine spectroscopic parameters for the \MgFexstate state, and we list the best fitted values in Table \ref{tab:exstateparams}. In our analysis, we fixed the ground state parameters to those determined by Anderson \textit{et al.}\cite{Anderson1994}, and reproduce these in Table \ref{tab:gdstateparams} for reference. The \MgFexstate state is well approximated by a Hund's case (a) coupling scheme for the angular momenta. Relevant details regarding the Hamiltonian are provided in Appendix \ref{Appendix:Hamiltonian}. The $\Lambda$-splitting is determined by the linear combination $p+2q$, and the hyperfine splittings are determined by the parameters $a$, $b_F+2c/3$ and $d$ \cite{Frosch1952a}\footnote[2]{\noindent Note that $b_F$ is related to $b$ and $c$ in equation 6.5 of Frosch and Foley \cite{Frosch1952a} by $b_F=b+c/3$.}. To independently measure $p$ and $q$, or $b_F$ and $c$, requires exciting to higher $J$ levels where the Hund's case (a) approximation breaks down. These parameter pairs are otherwise strongly correlated when fitted separately and so we state the linear combinations in Table \ref{tab:exstateparams}. The same reasoning applies for the parameters $A$ and $\gamma$.

\begin{table}
\caption{\label{tab:gdstateparams} Spectroscopic parameters of the \MgFgdstate state of MgF, reproduced from the work of Anderson \textit{et al.}\cite{Anderson1994}. In the original article, the hyperfine structure parameter of the fluorine nucleus $b(\textrm{F})=b_F(\textrm{F})-c(\textrm{F})/3$ was used.}
\begin{ruledtabular}
\begin{tabular}{ld}
Parameter &  \multicolumn{1}{c}{Value (MHz)}  \\
\hline
 $E_0(X^2\Sigma^+)$              &          0.0       \\
 $B              $ &      15496.8125    \\
 $     D         $ &          0.03238   \\
 $\gamma         $ &         50.697     \\
 $b_F(\textrm{F})         $&        214.2       \\
 $c(\textrm{F})           $ &        178.5       \\
\end{tabular}
\end{ruledtabular}
\end{table}

\begin{table}
\caption{\label{tab:exstateparams} Experimentally determined spectroscopic constants of the \MgFexstate state of MgF and their standard deviation (SD). The $\Lambda$-doubling parameters $p$ and $q$, and the spin-orbit ($A$) and spin rotation ($\gamma$) constants are strongly correlated. We state their linear combinations $p+2q$ and $A+\gamma$ that are well constrained by the fit.}
\begin{ruledtabular}
\begin{tabular}{lddd}
Parameter & \multicolumn{1}{c}{Value (MHz)}  & \multicolumn{1}{c}{SD} & \multicolumn{1}{c}{SD$\sqrt{Q}$\footnote{SD$\sqrt{Q}$ includes correlations
between parameters\cite{Watson1977}.}}\\
\hline
 $E_0(A^2\Pi)-\frac{1}{2}\gamma    $ &  
 834855315.2&         3 &       4\\
 $A+\gamma              $ &    1091346 &         3 &      3 \\
 $B              $ &      15788.2 &      0.3 &         0.4\\
 $\gamma         $ &        -53 &         16 &       136\\
 $p+2q           $ &         15 &         2 &       2\\
 $a($F$)           $ &        109 &          6 &         7\\
 $b_F($F$)+2c/3    $ &        -52 &         14 &        16 \\
 $d($F$)           $ &        135  &          7  &         7\\

\end{tabular}
\end{ruledtabular}
\end{table}

\subsection{Hyperfine structure}

\begin{figure*}
    \centering
    \includegraphics[]{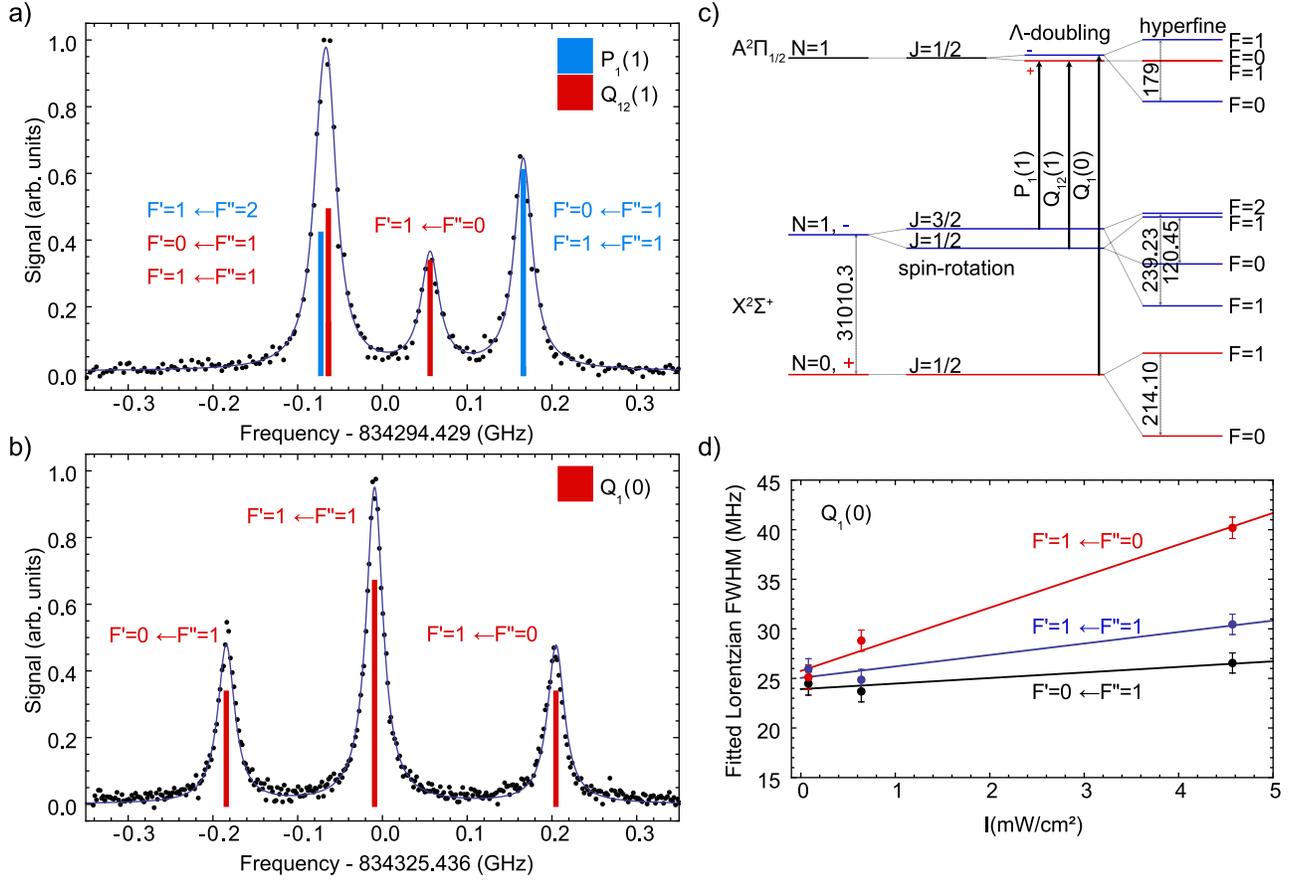}
    \caption{Hyperfine resolved spectra of the P$_1$/Q$_{12}$(1) (a) and Q$_1$(0) (b) lines. Experimental data are shown as black dots, and the line positions simulated from the fitted spectroscopic constants are shown as colored bars. The dark blue lines are Lorentzian fits to the experimental data to determine the line-center. (c) Energy level scheme of the hyperfine states involved in the transitions shown in (a) and (b). Energy differences are given in MHz. The two hyperfine states in the positive parity doublet in \MgFexstateLower, $J=1/2$ are not resolved; the calculation from spectroscopic constants determines the spacing to \SI{0.4}{\mega\hertz}. (d) Dependence of the linewidth of the three hyperfine lines of the Q$_1(0)$ spectrum as a function of the peak laser intensity. The hyperfine lines broaden differently with increasing intensity because of hyperfine and rotational pumping. Straight lines are linear fits to the data. The relative slopes agree well with simulations using rate equations.}
    \label{fig:levelschemeandspectra}
\end{figure*}

Fig. \ref{fig:levelschemeandspectra} a and b show hyperfine-resolved spectra of the P$_{1}$/Q$_{12}$(1) and Q$_{1}$(0) lines of the \mainMgFTransLower transition respectively. These lines originate from different rotational levels in the \MgFgdstate ground state and reach opposite parity levels in the same $J'=1/2$ level of the excited state. The structure of the P$_{1}$/Q$_{12}$(1) line is dominated by the ground state fine and hyperfine interactions, and the excited state hyperfine interaction is not resolved. However, in the negative parity $\Lambda$-doublet, the excited state splitting is \SI{179}{\mega\hertz}, and we resolve both the ground and excited state hyperfine structure in our spectra of the Q$_{1}$(0) line (Fig. \ref{fig:levelschemeandspectra} b). This is caused by a dependence of the magnetic hyperfine interaction on the sign of $\Lambda$, and therefore a difference in the linear combinations of $\Lambda$ states \cite{Frosch1952a}. The magnetic hyperfine constant $d($F$)$ encapsulates this effect, and its influence on the hyperfine splittings is illustrated in the level scheme of Fig. \ref{fig:levelschemeandspectra} c.\\

The hyperfine parameters $a,b_F,c$ and $d$ are related to properties of the electronic wavefunction in the molecule. A first-order approximation was initially described by Frosch and Foley\cite{Frosch1952a} and then subsequently simplified and corrected by Dousmanis\cite{Dousmanis1955}\footnote[3]{In particular, we note the correction to the value of $d$, acknowledged by Frosch and Foley.}. The interaction between the electron and nuclear magnetic moment can be split into two parts: one part is sensitive to the electron density at the nucleus, contained in $b_F$ and the other is sensitive to the orbital and spin wavefunction away from the nucleus, which determines $a, c$ and $d$. The coordinates of the electron relative to the interacting nucleus are expressed in the form $(r_1,\chi)$, where $r_1$ is the distance from the nucleus and $\chi$ is the opening angle subtended with respect to the internuclear axis. According to Dousmanis\footnote[4]{Here we use SI units, and our equations relate to the CGS units of Dousmanis by a factor $\mu_0/(4\pi)$.}, 

\begin{equation}
\label{eq:Dousmanis}
\begin{aligned}
 a  &= \frac{\mu_0}{2\pi}\mu_B\mu(\mathrm{F}) \expect{1/r_1^{3}},\\
 b_F  &= b + c/3 = \frac{4}{3}\mu_0\mu_B\mu(\mathrm{F})\psi^2(0),\\
 c &= \frac{3}{4\pi}\mu_0\mu_B\mu(\mathrm{F})\expect{\frac{3\cos^2\chi-1}{r_1^3}},\\
 d &= \frac{3}{4\pi}\mu_0\mu_B\mu(\mathrm{F})\expect{\sin^2\chi/r_1^3}
\end{aligned}
\end{equation}

\noindent Here, $\mu(\mathrm{F}) = 5.25\mu_N$ is the magnetic moment of the fluorine nucleus \cite{Stone2005}, $\mu_N$ is the nuclear magneton, and $\mu_B$ is the Bohr magneton. The angled brackets denote expectation values of the \MgFexstate electronic wavefunction, and $\psi^2(0)$ represents its probability density at the fluorine nucleus. From the value of $a$ we find a typical radius $\tilde{r} = \expect{r_1^{-3}}^{-1/3} = \SI{0.88}{\angstrom}$, roughly half the internuclear equilibrium separation of \SI{1.75}{\angstrom}. Approximating $\expect{\sin^2\chi/r_1^3}\approx \expect{\sin^2\chi}/\tilde{r}^3$ leads to a typical value $\tilde{\chi} = \ang{65.3}$. A wavefunction uniformly distributed over $\chi$ has $\tilde{\chi}=\ang{54.5}$. These observations suggest that, much like in the ground electronic state, there is appreciable electron density between the two nuclei.\\
This is in stark contrast with the other group II monofluorides CaF, SrF and BaF, where there is no resolvable interaction with the fluorine spin \cite{Bernath1980,Bernath1981,Kaendler1989,Chen2016a,Xu2017}, and upper limits of a few MHz have been inferred for $d($F$)$. It is only for the fermionic isotopologues of these molecules, where the metal nucleus has non-zero spin, that structure has been observed, and $d(^{87}\textrm{Sr})$, $d(^{135}\textrm{Ba})$, and $d(^{137}\textrm{Ba})$ could be determined \cite{Le2009,Steimle2011}.\\
Finally, we note that the hyperfine interaction causes mixing between the $X^2\Sigma^+(N=3,F=2)$ and $X^2\Sigma^+(N=1,F=2)$ states, and between the $A^2\Pi(J=1/2,F=1)$ and $A^2\Pi(J=3/2,F=1)$ states. This mixing results in losses from the P$_{11}/$Q$_{12}$(1) optical cycling transition. For $^{24}$MgF we calculate a branching ratio of $1.6\times 10^{-6}$ for $^{24}$MgF to the $N=3$ states. For $^{25}$MgF these losses are estimated to be an order of magnitude larger. A more detailed analysis of this effect is given in Appendix A.

\subsection{$\Lambda$-doubling}

In their discussion of the $\Lambda$-doubling in MgF, Walker and Richards derived values of $p$ and $q$ for MgF by extrapolation of the $\Lambda$-splitting observed at $J'+1/2 > 16$ \cite{Walker1970}. Their analysis gives $p+2q= \SI{-50.3}{MHz}$, which disagrees with our result in both the sign of the interaction and the magnitude. This discrepancy is likely due to the omission of the spin-rotation interaction in their analysis of the $^2\Sigma^+ - ^2\Sigma^+$ bands of MgF, and due to insufficient resolution in the A$^2\Pi \leftarrow \mathrm{X}^2\Sigma^+$ absorption spectra. We here provide an improved measurement and update their discussion of the origin of $\Lambda$-doubling in the group II monofluorides. In Table \ref{tab:vanVleckComp}, we compare measured values of $p + 2q$ with the values obtained according to Van Vleck's pure precession approximation \cite{VanVleck1929,Hinkley1972}. The approximation is valid when the $\Lambda$-doubling is dominated by the interaction with a single $^2\Sigma$ state, whose $\sigma$ molecular orbital is mainly derived from an atomic $p$ orbital \cite{Mulliken1931}. Under this assumption,
\begin{equation}
\begin{aligned}
    &p= p_{\textrm{vv}}= \frac{2ABl(l+1)}{E_{\Pi}-E_{\Sigma}} \hspace{0.3cm},\\
    &q= q_{\textrm{vv}} = \frac{2B^2l(l+1)}{E_{\Pi}-E_{\Sigma}} \hspace{0.3cm}.
    \label{eq:vv}
\end{aligned}
\end{equation}

\noindent Here, $l=1$ is the orbital angular momentum of the unpaired electron, and $E_{\Pi}-E_{\Sigma}$ is the energy difference between the interacting states. Equations \eqref{eq:vv} apply for the interaction with a $\Sigma^+$ state, with the sign reversing when the interaction is with a $\Sigma^-$ state. In the table we give $p_{\mathrm{vv}} + 2q_\mathrm{{vv}}$ values for pure precession with the nearby  B$^2\Sigma^+$ states which are higher in energy, and this appears to work well for the heavier monofluorides CaF, SrF and BaF. The trend is primarily due to the decreasing spin-orbit interaction moving up the group. In the case of MgF, the interaction changes sign and is an order of magnitude smaller than expected from the interaction with the B$^2\Sigma^+$ state. Therefore, $\Lambda$-doubling in MgF is more complex than for the heavier group II monofluorides and may comprise interactions with many $\Sigma$-states.
\textit{Ab initio} calculations by Kang \textit{et al.} \cite{Kang2015} imply that the B$^2\Sigma^+$ state of MgF is of mixed character, which may explain the marked difference in $\Lambda$-doubling compared to the heavier group II monofluorides.

\begin{table}
\caption{\label{tab:vanVleckComp} A comparison of the experimentally obtained $\Lambda$-doubling constants of the \MgFexstate state $p+2q$, with the value of $p_{\textrm{vv}}+2q_{\textrm{vv}}$ as discussed in the text, for the group II monofluorides. All values are stated in \SI{}{MHz}, and for MgF we use values obtained in this study. Sources of spectroscopic data are shown in the column headings where needed.}
\begin{ruledtabular}
\begin{tabular}{lcccc}
 & \multicolumn{1}{c}{MgF \cite{Barrow1967}} & \multicolumn{1}{c}{CaF \cite{Kaledin1999}} & \multicolumn{1}{c}{SrF \cite{Steimle1993,Steimle1977}} & \multicolumn{1}{c}{BaF \cite{Barrow1967a}} \\
\hline
   $10^{-3}(p+2q)$ (Exp.) &    \SI{0.015}{}$^a$  & \SI{-1.32}{}  & $\SI{-3.90}{}^b$  & $\SI{-7.67}{}^b$ \\
    $10^{-3}(p_{\textrm{vv}} + 2q_{\textrm{vv}})$ & \SI{-0.25}{}    &   \SI{-1.31}{} &   \SI{-4.17}{} &     \SI{-6.66}{} \\
\end{tabular}
\end{ruledtabular}
\footnotetext{This work.}
\footnotetext{In these studies, only $p$ is reported, and $q$ is assumed to be zero.}
\end{table}

\section{Radiative lifetime of the \MgFexstate,$v'=0$ level}
\label{sec:lifetime}
So far, the radiative lifetime of the \MgFexstate state is known only theoretically \cite{Pelegrini2005}. We determine the lifetime experimentally by measuring the Q$_1$(0) spectral line shape (Fig. \ref{fig:levelschemeandspectra} b) at low laser intensity. This line is convenient because the hyperfine structure in both ground and excited states is fully resolved, and the ground $N=0$ rotational level contains the largest population of slow molecules for which the residual Doppler broadening is smallest. For these measurements, we reduced the expected Doppler broadening to \SI{1}{MHz} by replacing the 2x\SI{2}{\square\milli\meter} square molecular beam aperture in the detector with a \SI{1}{mm} slit. We measure spectra at several probe laser intensities, and extract the full width at half-maximum (FWHM) for each line with a Lorentzian fitting function. Fig. \ref{fig:levelschemeandspectra} d plots the linewidths for the three hyperfine components of the Q$_1$(0) line against the peak laser intensity together with linear fits (solid lines). Linear fits to the data show that each line broadens differently with increasing laser intensity. This broadening occurs at laser intensities well below the predicted two-level saturation intensity, $I_s=\pi h c\Gamma/3\lambda^3=$ \SI{62}{\milli\watt\per\centi\meter\squared}, and is the result of optical pumping between rotational and hyperfine states of the molecule, which we discuss in the following paragraph. 

In general, the spectrum of an open transition will broaden when the number of photon scattering events is sufficient to optically pump the molecule to a state not addressed by the laser. This broadening can occur at an arbitrarily low laser intensity $I$, provided that the interaction time $t_i$ is large enough. In the absence of hyperfine structure, molecules are pumped to $N''=2$ on the Q$_{1}(0)$ line after an average of three scattering events. With the inclusion of hyperfine structure, molecules are pumped to both $N''=2$ and also between hyperfine levels of $N''=0$, further reducing the number of scattering events. For our experiments where the typical laser interaction time is $t_i=\SI{10}{\micro\second}$, this effect becomes significant even when $I\sim 10^{-3}I_{s}$. To verify our understanding, we simulate the interaction with the laser using rate equations, the measured hyperfine splittings, and the branching ratios for each hyperfine decay channel. We assume all three polarization components are excited with equal probability; this is a reasonable approximation given the magnetic field in the detector mixes the ground states by spin precession during the interaction time with the laser. We fit the simulated spectra with the same Lorentzian model, and find that the model predicts the relative broadening rates within the experimental uncertainties. A complete quantitative treatment of this effect requires detailed information about the laser profile and the collection optics, which is beyond the requirements of this paper. We refer the interested reader to the work of Wall \textit{et al.}\cite{Wall2008} as an example.  

To estimate an upper bound for the true Lorentzian linewidth $\Gamma/(2\pi)$, where $\Gamma=1/\tau_0$ and $\tau_0$ is the radiative lifetime of the \MgFexstate,$v'=0$ level, we use the spectrum taken at a peak intensity of \SI{80}{\micro\watt\per\centi\meter\squared}, shown in Fig. \ref{fig:levelschemeandspectra} b. Here the FWHM of the individual hyperfine lines are consistent within the uncertainties, indicating that the effect of optical pumping is small. We fit the data to a sum of three Voigt profiles, finding a Lorentzian FWHM of $\Gamma/(2\pi)=\SI{22.0(5)}{MHz}$ and a Gaussian FWHM of $3.8(1.8)$\SI{}{MHz}. The ground state splitting can be estimated from our experimental spectra, and compared to the precise measurements of Anderson \textit{et al.} to estimate an uncertainty on the linearity of the laser scan. We find agreement within $\pm \SI{1}{MHz}$, consistent with the Yb measurements presented in Section \ref{experimental}. The Gaussian contribution to the lineshape arises from the Doppler effect, residual Zeeman shifts and laser frequency instability. To estimate its systematic uncertainty, we measured the $(3s^27s) ^2S_{1/2} \leftarrow (3s^23p) ^2P_{1/2}$ narrow transition in a buffer gas beam of atomic aluminum. The atoms are produced in the same beam machine, with a forward velocity similar to the MgF molecules. The probe-light at \SI{225.8}{nm} is generated from the fourth harmonic of the same titanium sapphire laser, which increases the laser frequency noise by at least a factor two. The sensitivity to Doppler shifts arising from the shorter probe wavelength is increased by a factor 1.6 relative to the Q$_{1}(0)$ line in MgF. The Al transition has a natural linewidth of \SI{2.5}{\mega\hertz} \cite{Paschen1932}, and using this value we arrive at an upper bound of the Gaussian FWHM contribution of \SI{9.2(2)}{\mega\hertz} at the Al detection wavelength. We therefore vary the Gaussian contribution in the final fit between 1 and \SI{4.6}{\mega\hertz}, and find that this changes the fitted value of $\Gamma/(2\pi)$ by at most \SI{1}{MHz}. From this, we estimate a lower bound for the radiative lifetime of the \MgFexstate,$v'=0$ state to be $\tau_0 = 1/\Gamma = (7.23\pm0.16_{stat}\pm0.33_{sys})$ ns, in agreement with the theoretical prediction by Pelegrini \textit{et al.} \cite{Pelegrini2005} of \SI{7.16}{\nano\second}. The uncertainty is dominated by the systematic uncertainty in the Gaussian contribution to the spectral lineshape.   
\vspace{-2mm}
\section{Electric dipole moment measurements}

The application of an external electric field $\boldsymbol{E}$ in the detector introduces an additional term,

\begin{equation}
    H_{Stark}=-\boldsymbol{\mu}_i\cdot\boldsymbol{E},
    \label{eqn:HStark}
\end{equation}

\noindent to the Hamiltonian given in Equation \eqref{eq:hamil}. Here, $\boldsymbol{\mu}_i$ is the vector dipole moment operator in electronic state $i$. From the Stark splitting and shifting of the LIF spectra we can determine the magnitude of the dipole moments $|\boldsymbol{\mu}_A|$ and $|\boldsymbol{\mu}_X|$.

The Stark Hamiltonian \eqref{eqn:HStark} couples nearby molecular states of opposite parity having the same total angular momentum projection $M_F$ onto an axis parallel to $\boldsymbol{E}$. To calculate the energies under an applied field, we diagonalize the Hamiltonian matrix constructed using the relevant Hund's case basis functions, including all levels with $J\leq 9/2$. In the electronic ground state the Stark Hamiltonian mainly mixes states separated by one unit of the quantum number $N''$. For the electric fields applied in this study of up to \SI{10.6}{kV\per\centi\meter}, the Stark shift is quadratic. For the excited states, the dominant interaction is between the closely spaced $\Lambda$-doublet levels. The interaction is strong compared to this splitting and results in a linear Stark shift even for low electric field strengths. The Stark effect overcomes the hyperfine interaction at modest electric field strengths of about \SI{0.1}{kV\per\centi\meter}, after which the levels separate by their value of the angular momentum projection $M_{J'}$. 

\begin{figure*}
    \centering
    \includegraphics[]{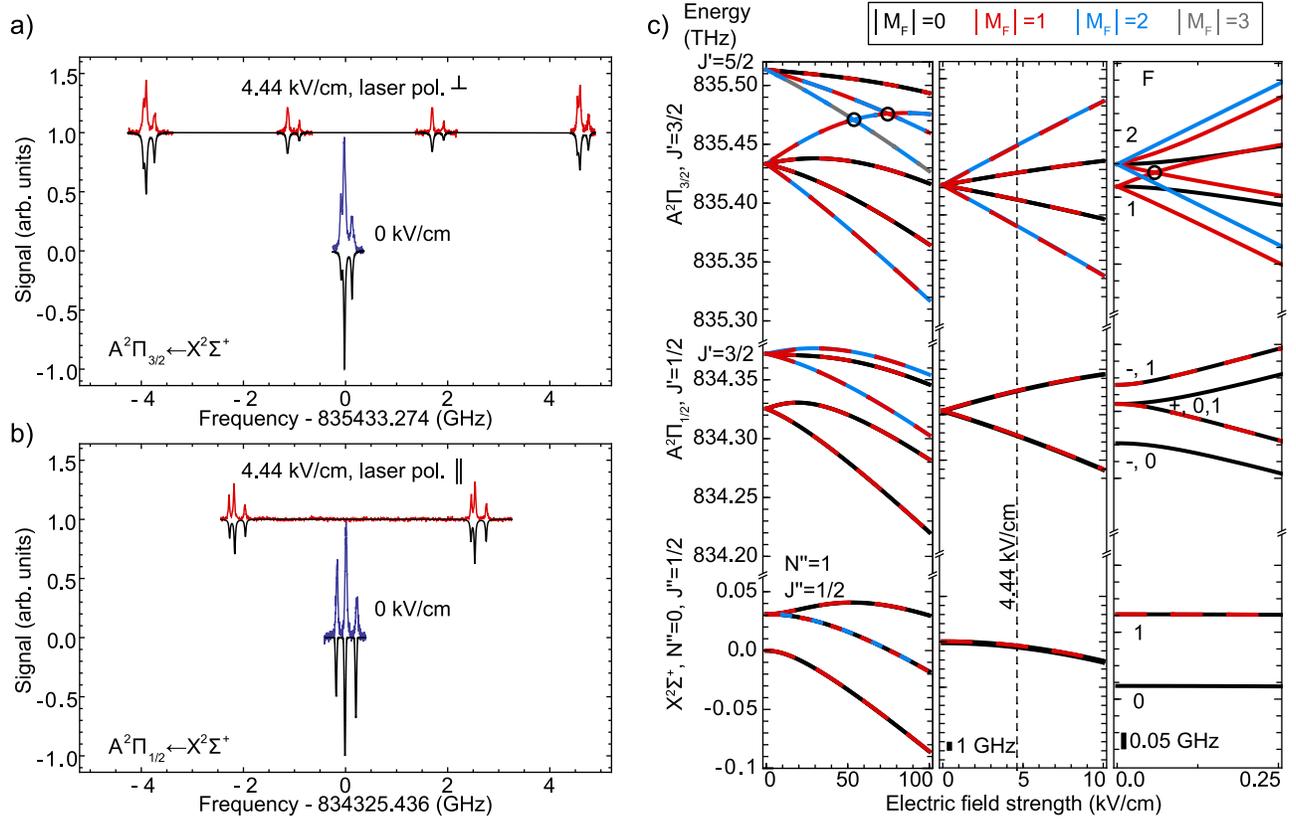}
   \caption{Stark effect measurement of the R$_{21}$(0) (a) and Q$_{1}$(0) (b) lines of the \mainMgFTrans transition in MgF. The field-free spectra are shown in blue and when an electric field of 4.44 kV/cm is applied in red. The simulated spectra assuming $\mu_X=2.88(20)$ D and $\mu_A=3.20(22)$ D for the ground and excited state dipole moments, respectively, are shown in black. (c) Calculated Stark shifts of the $X^2\Sigma^+, N''=0, J''=1/2$, $A^2\Pi_{1/2}, J'=1/2$, and $A^2\Pi_{3/2}, J'=3/2$ levels. The left panel shows the shifts of up to \SI{100}{\kilo\volt\per\centi\meter}, the center shows the region of $0-$\SI{10}{\kilo\volt\per\centi\meter} and the right panel shows a zoom-in to $0-$\SI{0.25}{\kilo\volt\per\centi\meter}. Avoided crossings are marked with black circles. The different $|M_F|$ components are color-coded. If the Stark curves of two states with the same $|M_F|$ overlap at the resolution of the figure, they are shown as dashed lines. The electric field used in the spectra (a and b) is indicated with a dashed vertical line.}
    \label{fig:starkshifts}
\end{figure*}

Fig. \ref{fig:starkshifts} a and b show spectra of the  R$_{21}$(0) and Q$_{1}$(0) lines, respectively, under field-free conditions and with an electric field of \SI{4.44}{\kilo\volt\per\centi\meter} applied in the detector. The total span of each spectrum is determined by the Stark effect of the excited state, and the small shift in the gravity center is due to the ground state Stark shift. With the laser polarization oriented perpendicular to the electric field, we excite and observe all four $M_{J'}$ components of the R$_{21}$(0) lines in Fig. \ref{fig:starkshifts} a. We measure spectra at various applied fields and fit the modeled spectra to extract best fit values for the dipole moments, obtaining $\mu_A=3.20\pm0.01_{stat}\pm0.22_{sys}$~D and $\mu_X=2.88\pm0.03_{stat}\pm0.20_{sys}$~D. The systematic uncertainty is dominated by the determination of the electric field between the mesh electrodes. We assume a measurement uncertainty of \SI{0.3}{mm} in the mesh separation of \SI{9}{\milli\meter}, and with finite element modelling we estimate a reduction of $2\%$ in the electric field strength relative to infinite plate electrodes of the same separation. The combination of these effects far exceeds the statistical uncertainty from the fitting procedure. Simulated spectra using the best fit values are shown inverted in Fig. \ref{fig:starkshifts} a and b, demonstrating good quantitative agreement with the measurements. The FWHM of the lines measured at low and high field are consistent within \SI{1}{MHz}, from which we deduce that spatial inhomogeneity of the electric field across the probe beam is below $0.1\%$. 

To examine the behavior of the energy levels in MgF in electric fields in more detail, we show the simulated Stark shifts for low-$J$ levels at high, moderate and small electric field strengths in Fig. \ref{fig:starkshifts} c. Between 10 and \SI{100}{kV/cm} (left panel), different rotational levels interact significantly; this leads to avoided crossings between the excited states, indicated by circles in the Fig.\ref{fig:starkshifts}. At intermediate fields up to \SI{10}{kV/cm} (center panel), the Stark effect of the ground state becomes comparable to the excited state. At electric fields below \SI{0.25}{kV/cm} (right panel), mixing in the ground state is negligible, whereas in the excited state the nearby $\Lambda$-states mix significantly.

The \MgFgdstate ground state dipole moment has been predicted theoretically  \cite{Langhoff1986,Fowler1991,Kobus2000,Pelegrini2005,Wu2015,Hou2017}. The values obtained by different methods range from \SI{2.67}{D}\cite{Pelegrini2005} to \SI{3.126}{D}\cite{Wu2015}. We note that our measured value of $\mu_X$ is in good agreement with the theoretical value obtained by Fowler and Sadlej\cite{Fowler1991}, $\mu_X =\SI{2.8611}{D}$, using the complete active space self-consistent field (CASSCF) method. Pelegrini \textit{et al.} also calculated $\mu_A=4.23$ D, which is in poor agreement with our experimental results. Table \ref{tab:dipolemoments} compares our experimental values to those of other group II monofluorides. Steimle \textit{et al.}\cite{Steimle2011} reported two different values of the electric dipole moment for the \MgFexstateLower and \MgFexstateUpper states of BaF, which can be explained by interactions with other states. Within the  statistical uncertainty of 0.3\%, our Stark shifted spectra of MgF are well described by a single dipole moment for the \MgFexstate state. This suggests very little influence of nearby perturbing states on the dipole moment, in contrast to BaF.

\section{Electric field induced rotational branching}
The laser cooling scheme for MgF relies on the parity ($P$) and angular momentum ($J$) selection rules of electric dipole transitions. In zero electric field, optical cycling is possible on the P$_1/$Q$_{12}(1)$ transition, which excites molecules from the $N''=1$ ground states to the $J'=1/2, P'=+$ excited states.
%
However, a small electric field $E$ results in an exchange of population, $\epsilon^2$, between excited states of opposite parity. The relative transition strength from $N''=1$ becomes  $1-\epsilon^2$ when exciting the states of mostly positive character, and these states decay to $N''=0,2$ with probability $\epsilon^2$. In addition, excitation to the states with mostly negative parity character becomes weakly allowed, with a relative transition strength $\epsilon^2$. These states decay to $N''=0,2$ with near unit probability. This second loss channel cannot be neglected, because efficient optical cycling requires frequency sidebands to address the ground state hyperfine splitting, causing the weak transitions to be driven near resonance. As a result, we estimate the loss probability from the cooling cycle due to uncontrolled electric fields, to lowest order in $\epsilon$, as $\mathcal{P}_{\mathrm{loss}}\approx 2\epsilon^2$.

\begin{figure}
    \centering
    \includegraphics[]{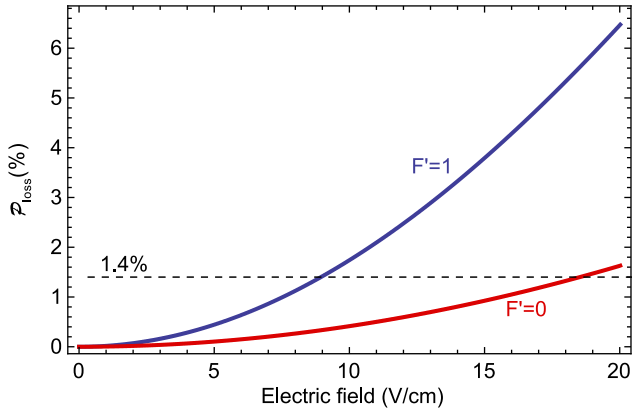}
   \caption{Calculated population mixing of opposite parity levels, $\epsilon^2$, for the different hyperfine levels of the \MgFexstate, $J'=1/2$ states in an electric field. This results in unwanted rotational branching to $N''=0$ and 2 in the ground state when driving the P$_1$/Q$_{12}(1)$ laser cooling transition. The $\left|M_{F'}\right|=0,1$ levels in $F'=1$ are shown in blue, the $F'=0, \left|M_{F'}\right|=0$ level is shown in red. The dashed line indicates the level of expected losses due to vibrational branching to $v''=1$. The quadratic loss coefficients are $\alpha_0=4\times10^{-5}\SI{}{cm^2/V^2}$ and $\alpha_1=2\times10^{-4}\SI{}{cm^2/V^2}$.}
    \label{fig:lossrate}
\end{figure}

In Fig. \ref{fig:lossrate}, we plot $\mathcal{P}_{\mathrm{loss}}$ versus the electric field strength for the two possible values of $F'$, using our measured spectroscopic constants. We fit the data below 5 V/cm to $\mathcal{P}_{\mathrm{loss}}=\alpha_{F'}|E|^2$, finding that $\alpha_0=4\times10^{-5}\SI{}{cm^2/V^2}$ and $\alpha_1=2\times10^{-4}\SI{}{cm^2/V^2}$. For the P$_1$/Q$_{12}$(1) cooling transition, electric fields of \SI{18}{\volt\per\centi\meter} and \SI{9}{\volt\per\centi\meter} lead to a mixing of the parity eigenstates of about \SI{1.4}{\%}, for $F'=0$ and $1$, respectively. At this level, losses from the optical cycle due to parity mixing match losses to the $v''=1$ vibrational manifold in the \MgFgdstate state predicted by Ref. \cite{Pelegrini2005}. We note that such electric field induced losses have already been observed experimentally in SrF \cite{Norrgard2016} and AlF \cite{Hofsass2021}.  
\begin{table}
\caption{\label{tab:dipolemoments} Reported dipole moments of group-II monofluorides, in debye. Values for MgF are from this study.}
\begin{ruledtabular}
\begin{tabular}{lcccc}
Parameter & MgF & CaF & SrF & BaF  \\
\hline
 $\mu(X^2\Sigma^+)$              &          2.88(20)     & 3.07(7)\cite{Childs1984} &  3.4963(4)\cite{Ernst1985} & 3.170(3)\cite{Ernst1986} \\
  $\mu(A^2\Pi_{1/2})$              &          3.20(22)     &  &  & 1.50(2)\cite{Steimle2011} \\
   $\mu(A^2\Pi_{3/2})$              &          3.20(22)    & 2.45(6)\cite{Ernst1989} & 2.064(50)\cite{Kaendler1989} & 1.31(2)\cite{Steimle2011}  \\

\end{tabular}
\end{ruledtabular}
\end{table}

\section{Conclusion}
We recorded hyperfine-resolved CW LIF spectra of 25 low-$J$ lines of the \mainMgFTrans transition in MgF, and analyzed the $^{24}$MgF isotopologue in detail. By fitting the eigenvalues of the effective Hamiltonian to the measured line positions, we determined the spectroscopic parameters of the \MgFexstate state that are relevant for laser cooling experiments: rotational, fine- and hyperfine structure constants. We calibrated our wavemeter using the precisely known Yb $(6s6p)^1P_1\leftarrow(6s^2)^1S_0$ transition frequencies \cite{Kleinert2016}, and correct a \SI{-4.1}{GHz} systematic error in the line frequencies presented by Xu \textit{et al.} \cite{Xu2019}. Transition isotope shifts between the $^{24}$MgF and $^{26}$MgF isotopologues were recorded, and we observe an unexplained \SI{470}{MHz} transition frequency shift which may indicate deviations from the Born-Oppenheimer approximation. We studied the broadening of the hyperfine lines due to optical pumping and recorded high-resolution spectra of the Q$_1(0)$ line to determine the radiative lifetime of the \MgFexstate, $v'=0$ level to be $\tau_0=\SI{7.23(36)}{ns}$. 

By studying the fluorescence spectra under an applied electric field, we experimentally determined the dipole moments of the \MgFgdstate and \MgFexstate states. Our value for the ground state, $\mu_X=2.88\pm0.03_{stat}\pm0.20_{sys}$~D is in good agreement with the value predicted by \textit{ab initio} calculations \cite{Kobus2000,Langhoff1986,Fowler1991,Wu2015,Hou2017}. Using our value of  $\mu_A=3.20\pm0.01_{stat}\pm0.22_{sys}$~D, we predict the electric field strength at which parity-mixing in the excited states limits optical cycling on the P$_1$/Q$_{12}$(1) line. We find that \SI{9}{\volt\per\centi\meter} is sufficient for unwanted rotational branching to match the expected vibrational branching. To scatter more than $10^4$ photons, stray electric fields have to be controlled to below the \SI{1}{\volt\per\centi\meter} level. Coincidentally, the hyperfine structure in the $J'=1/2,P'=+1$ level is less than \SI{1}{\mega\hertz}, which simplifies the laser cooling scheme significantly. On the contrary, the hyperfine splitting in the $J'=1/2,P'=-1$, level is large, which increases the separation between opposite parity hyperfine levels. This reduces the sensitivity of the optical cycling scheme to stray electric fields substantially.  

There are a number of notable differences between MgF and the heavier group II monofluorides CaF, SrF and BaF. First, the sign of the $\Lambda$-splitting is inverted in MgF, and its magnitude is about 100 times smaller. Second, MgF has the largest dipole moment in the \MgFexstate state, whereas it has the smallest dipole moment in its ground state. Third, the interaction of the electronic angular momentum with the fluorine nuclear spin leads to a resolvable hyperfine structure in the excited state of MgF, and from our measured hyperfine constants we can infer that the electronic wavefunction has significant probability density between the nuclei. This supports the conclusions of Anderson \textit{et al.} \cite{Anderson1994} regarding the greater covalency of the chemical bonding in MgF. 

Finally, we note that our measurements form a stringent set of benchmarks for precise quantum chemical calculations on MgF. The hyperfine constants and dipole moment measurements are strong benchmarks for molecular orbital calculations of the \MgFgdstate and \MgFexstate states, while the transition isotope shifts presented constrain vibrational constants and deviations from the Born-Oppenheimer approximation. 

\section*{Acknowledgments} \noindent The authors acknowledge the mechanical and electronic workshops of the Fritz Haber Institute for their expert technical assistance. This project received funding from the European Research Council (ERC) under the European Union’s Horizon 2020 Research and Innovation Programme (CoMoFun, Grant Agreement No. 949119).

\section*{Conflict of Interest}
\noindent The authors have no conflicts to disclose.
\section*{Data Availability}
\noindent The data that support the findings of this study are openly available in Zenodo.com at https://doi.org/10.5281/zenodo.6110184, and may be used under the Creative Commons Attribution 4.0 International license.

\appendix
{\section{Hamiltonian}
\label{Appendix:Hamiltonian}
To understand the influence of the different spectroscopic parameters in the Hamiltonian \eqref{eq:hamil}, we discuss subspaces of the Hund's case (a) basis $|\Omega,J,P,F\rangle$ relevant for the excited A$^2\Pi$ states. Mixing between $|\Omega,J,P,F\rangle$ states is restricted to be within subspaces with the same $P$ and $F$ due to parity and angular momentum selection rules. The pure spin-orbit and rotational Hamiltonian, $H_{\Omega_i,\Omega_k}$, has diagonal elements, 
\begin{equation}
    H_{\Omega,\Omega}(J)=(\Omega-1)A+B\left[J(J+1)-2\Omega+5/4\right]+\left(\Omega-{3}/{2}\right)\gamma
\end{equation}
\noindent and off-diagonals,     
\begin{equation}
    H_{\frac{1}{2},\frac{3}{2}}(J) =  H_{\frac{3}{2},\frac{1}{2}}(J) =-\left(B-\gamma/2\right)\sqrt{J(J+1)-3/4} \hspace{0.5cm} .
\end{equation}
\noindent Each $F=0,P$ subspace contains a single state and these do not mix. There are three states in a $|\Omega,J,P=\pm,F=1\rangle$ subspace, and we order these states as $|\Omega,J\rangle=|1/2,1/2\rangle$, $|1/2,3/2\rangle$, $|3/2,3/2\rangle$. The combined Hamiltonian $H(F=1,P=\pm)$, which includes the hyperfine and $\Lambda$-doubling interactions, is given by,
\begin{widetext}
\begin{equation}
\left(
\begin{array}{ccc}
H_{\frac{1}{2},\frac{1}{2}}\left(\frac{1}{2}\right) \mp\frac{\tilde{p}}{2}+\frac{a}{6}-\frac{\tilde{b}}{12} \mp\frac{d}{6} & h.c. & h.c. \\
-\frac{\sqrt{2} a}{3}+ \frac{\tilde{b}}{3 \sqrt{2}}\mp\frac{d}{3 \sqrt{2}} & H_{\frac{1}{2},\frac{1}{2}}\left(\frac{3}{2}\right)\pm\tilde{p}-\frac{a}{6}+\frac{\tilde{b}}{12} \mp\frac{d}{3} & h.c. \\
 \frac{\overset{\approx }{b}}{\sqrt{6}} & H_{\frac{1}{2},\frac{3}{2}}(\frac{3}{2})-\sqrt{3} q -\frac{\overset{\approx }{b}}{2 \sqrt{3}}& H_{\frac{3}{2},\frac{3}{2}}\left(\frac{3}{2}\right)-\frac{a}{2}-\frac{\tilde{b}}{4} \\
\end{array}
\right),
\label{eqn:HamFOne}
\end{equation}
\end{widetext}
Here, $\tilde{b} = b_F+2c/3$, $\overset{\approx}{b}=b_F-c/3$, $\tilde{p} = p +2q$, and $h.c.$ refers to the Hermitian conjugate. By applying perturbation theory to the Hamiltonian \eqref{eqn:HamFOne}, mixing of the $\ket{\Omega=1/2,J=1/2}$ and $\ket{\Omega=1/2,J=3/2}$ states due to the hyperfine interaction can be calculated. To second order, the population mixing $\zeta$ is, 
\begin{equation}
   \zeta= 2\left(\frac{\tilde{b}-d-2a}{18B -2a-d+\tilde{b}+9\tilde{p}} \right)^2 \approx \frac{1}{2}\left(\frac{1}{3}\frac{\tilde{b}-2a-d}{3B}\right)^2 
\end{equation}
The resulting loss from the cycling transition due to hyperfine mixing in the excited state, $\eta_A$ is,

\begin{equation}
    \eta_A = \frac{S(\mathrm{P}_{12})}{S(\mathrm{P}_{12})+S(\mathrm{R}_{12})+S(\mathrm{Q}_{1}) } \zeta
\end{equation}

\noindent Here, $S(\mathcal{L})$ is the H\"{o}nl-London factor for decay from $\Omega=1/2,J=3/2$ by transition $\mathcal{L}$, and we note that only the P$_{12}$ path results in loss from the $N=1$ ground states. Using our spectroscopic parameters, the loss probability from $F=1$ is $\eta_A=1.2\times 10^{-6}$. Loss can also occur via mixing with the $\ket{\Omega=3/2,J=3/2}$ states, but this is suppressed by about a factor $\left(A/B\right)^2$ and we neglect this term. A similar calculation for the ground state can be used to calculate the hyperfine mixing between $N''=1$ and $N''=3$, and results in a total loss $\eta_X\approx 0.4\times 10^{-6}$ from the cooling cycle. Therefore, the total loss probability from the $\ket{\Omega = 1/2,J=1/2,P=+1,F=1}$ excited states from hyperfine mixing is $\eta = 1.6\times 10^{-6}$.

To complete our discussion, we consider now the subspaces with $F\geq 2$, each of which contains four states. For $F=2, P=\pm$, the combined Hamiltonian matrix is 
\begin{widetext}
{\small{
\begin{equation}\label{eqn:HamFTwo}
\left(
\begin{array}{cccc}
 H_{\frac{1}{2},\frac{1}{2}}\left(\frac{3}{2}\right)-\frac{\tilde{b}}{20}\pm\tilde{p}+\frac{a}{10} \pm\frac{d}{5} & h.c. & h.c.& h.c. \\
  \frac{1}{5} \sqrt{\frac{3}{2}} \tilde{b}-\frac{\sqrt{6} a}{5}\pm\frac{1}{5} \sqrt{\frac{3}{2}} d & H_{\frac{1}{2},\frac{1}{2}}\left(\frac{5}{2}\right)+\frac{\tilde{b}}{20}\mp\frac{3 \tilde{p}}{2}-\frac{a}{10} \pm\frac{3 d}{10} & h.c. & h.c. \\
H_{\frac{1}{2},\frac{3}{2}}(\frac{3}{2})+\frac{\sqrt{3} \overset{\approx }{b}}{10}\mp\sqrt{3} q & -\frac{\overset{\approx }{b}}{5 \sqrt{2}} &  H_{\frac{3}{2},\frac{3}{2}}(\frac{3}{2})+\frac{3 \tilde{b}}{20}+\frac{3 a}{10} & h.c.\\
 \frac{\sqrt{3} \overset{\approx }{b}}{5} & H_{\frac{1}{2},\frac{3}{2}}(\frac{5}{2}) -\frac{\sqrt{2} \overset{\approx }{b}}{5}\pm3 \sqrt{2} q & -\frac{\tilde{b}}{5}-\frac{2 a}{5} & H_{\frac{3}{2},\frac{3}{2}}\left(\frac{5}{2}\right) -\frac{3 \tilde{b}}{20}-\frac{3 a}{10} \\
\end{array}
\right),
\end{equation}
}}
\end{widetext}
\noindent where the ordering of states is $|\Omega,J\rangle=|1/2,3/2\rangle$, $|1/2,5/2\rangle$, $|3/2,3/2\rangle$, $|3/2,5/2\rangle$. The eigenenergies of equation \eqref{eqn:HamFOne} and \eqref{eqn:HamFTwo} are, to first order, the diagonals of the matrices, and can be used to accurately determine $a,\tilde{b},d,\tilde{p}$. The parameters $\overset{\approx }{b}$ and $q$ contribute only as a second order correction to the energies, and appear as terms $\mathcal{O}\left(\frac{\overset{\approx }{b}^2}{A},\frac{q^2}{A},\frac{B}{A}\overset{\approx }{b}, \frac{B}{A}q\right)$ or smaller. As a result, an experiment just capable of resolving $\tilde{b}$ or $\tilde{p}$ would require about two orders of magnitude higher resolution to resolve similar values for $\overset{\approx }{b}$ and $q$, for low-$J$ lines. The off-diagonal matrix elements $H_{\Omega_i,\Omega_k}$ increase approximately linearly with $J$, which magnifies the role of the $q$ and $ \overset{\approx }{b}=b_F-c/3$ parameters as $J$ increases. The role of all 6 parameters will be significant in spectra of high-$J$ lines of MgF.}

\section{Observed line frequencies}
\LTcapwidth=0.48\textwidth
\begin{longtable}{llccccccc}
\caption{\label{tab:linelist}Fitted line positions used for the determination of the spectroscopic constants of the \MgFexstate state, presented in Table \ref{tab:exstateparams}. We show the observed line frequencies, the deviation observed-calculated (o-c) and their assignment. The presence of O- and S-lines with $\Delta J=2$ is a result of the hyperfine interaction in the ground state, which mixes different $J$-states and breaks the $(\Delta J=0,1)$ selection rule. The statistical uncertainty on the individual hyperfine lines is below \SI{1}{\mega\hertz}. On different days, the gravity center of a rotational line is reproducible to within 10 MHz.}
\endfirsthead
\hline
\hline
Label & Observed (MHz) & o-c & $\Omega$ & $J'$ & $F'$ & $N''$ & $J''$ & $F''$ \\
\hline
P$_{12}(3)$ & 834186158 & 7 & 1/2 & 3/2 & 1 & 3 & 5/2 & 2 \\
& 834186384 & 6 & 1/2 & 3/2 & 2 & 3 & 5/2 & 3 \\
O$_{1}(3)$ & 834186100 & 3 & 1/2 & 3/2 & 2 & 3 & 7/2 & 3 \\
\hline
P$_1(1)$ & 834294356 & 1 & 1/2 & 1/2 & 1 & 1 & 3/2 & 2 \\
& 834294595 & 1 & 1/2 & 1/2 & 1 & 1 & 3/2 & 1 \\
& 834294595 & 0 & 1/2 & 1/2 & 0 & 1 & 3/2 & 1 \\
Q$_{12}(1)$ & 834294485 & 0 & 1/2 & 1/2 & 1 & 1 & 1/2 & 0 \\

P$_{1}(2)$ & 834278982 & 0 & 1/2 & 3/2 & 2 & 2 & 5/2 & 3 \\
& 834279000 & -3 & 1/2 & 3/2 & 2 & 2 & 5/2 & 2 \\
& 834279038 & -2 & 1/2 & 3/2 & 1 & 2 & 5/2 & 2 \\
Q$_{12}(2)$ & 834279126 & -3 & 1/2 & 3/2 & 2 & 2 & 3/2 & 1 \\
& 834279167 & 0 & 1/2 & 3/2 & 1 & 2 & 3/2 & 1 \\
& 834279257 & 3 & 1/2 & 3/2 & 2 & 2 & 3/2 & 2 \\
\hline
Q$_1(0)$ & 834325252 & 3 & 1/2 & 1/2 & 0 & 0 & 1/2 & 1 \\
& 834325426 & -4 & 1/2 & 1/2 & 1 & 0 & 1/2 & 1 \\
& 834325639 & -5 & 1/2 & 1/2 & 1 & 0 & 1/2 & 0 \\
Q$_1(1)$ & 834341073 & -4 & 1/2 & 3/2 & 2 & 1 & 3/2 & 2 \\
& 834341204 & -6 & 1/2 & 3/2 & 1 & 1 & 3/2 & 1 \\
& 834341319 & 3 & 1/2 & 3/2 & 2 & 1 & 3/2 & 1 \\
R$_{12}(1)$ & 834340972 & -8 & 1/2 & 3/2 & 1 & 1 & 1/2 & 1 \\
Q$_{1}(3)$ & 834372660 & -2 & 1/2 & 7/2 & 3 & 3 & 7/2 & 3 \\
& 834372717 & 3 & 1/2 & 7/2 & 4 & 3 & 7/2 & 4 \\
R$_{12}(3)$ & 834372822 & -1 & 1/2 & 7/2 & 3 & 3 & 5/2 & 2 \\
& 834372939 & -4 & 1/2 & 7/2 & 3 & 3 & 5/2 & 3 \\
& 834373029 & 4 & 1/2 & 7/2 & 4 & 3 & 5/2 & 3 \\
\hline

R$_1(0)$ & 834372011 & 8 & 1/2 & 3/2 & 2 & 0 & 1/2 & 1 \\
& 834372047 & 6 & 1/2 & 3/2 & 1 & 0 & 1/2 & 1 \\
& 834372256 & 1 & 1/2 & 3/2 & 1 & 0 & 1/2 & 0 \\
R$_1(1)$ & 834418742 & 1 & 1/2 & 5/2 & 3 & 1 & 3/2 & 2 \\
& 834419025 & -4 & 1/2 & 5/2 & 2 & 1 & 3/2 & 1 \\
S$_{12}(1)$ & 834418793 & -6 & 1/2 & 5/2 & 2 & 1 & 1/2 & 1 \\
R$_1(2)$ & 834465601 & 5 & 1/2 & 7/2 & 4 & 2 & 5/2 & 3 \\
S$_{12}(2)$ & 834465923 & -1 & 1/2 & 7/2 & 3 & 2 & 3/2 & 2 \\
& 834465670 & -2 & 1/2 & 7/2 & 3 & 2 & 5/2 & 2 \\
\hline
\hline
Q$_{2}(2)$ & 835340294 & -8 & 3/2 & 3/2 & 1 & 2 & 3/2 & 1 \\
& 835340370 & 1 & 3/2 & 3/2 & 2 & 2 & 3/2 & 1 \\
& 835340428 & 1 & 3/2 & 3/2 & 1 & 2 & 3/2 & 2 \\
& 835340494 & 0 & 3/2 & 3/2 & 2 & 2 & 3/2 & 2 \\
P$_{21}(2)$ & 835340215 & -6 & 3/2 & 3/2 & 2 & 2 & 5/2 & 3 \\
& 835340234 & -9 & 3/2 & 3/2 & 2 & 2 & 5/2 & 2 \\
\hline
R$_{2}(1)$ & 835402172 & -1 & 3/2 & 3/2 & 1 & 1 & 1/2 & 1 \\
& 835402247 & 7 & 3/2 & 3/2 & 2 & 1 & 1/2 & 1 \\
& 835402296 & 3 & 3/2 & 3/2 & 1 & 1 & 1/2 & 0 \\
Q$_{21}(1)$ & 835402162 & -2 & 3/2 & 3/2 & 1 & 1 & 3/2 & 2 \\
& 835402236 & 5 & 3/2 & 3/2 & 2 & 1 & 3/2 & 2 \\
& 835402405 & 2 & 3/2 & 3/2 & 1 & 1 & 3/2 & 1 \\
& 835402471 & 0 & 3/2 & 3/2 & 2 & 1 & 3/2 & 1 \\

R$_{2}(3)$ & 835439653 & 7 & 3/2 & 7/2 & 3 & 3 & 5/2 & 2 \\
& 835439771 & 5 & 3/2 & 7/2 & 3 & 3 & 5/2 & 3 \\
& 835439802 & 1 & 3/2 & 7/2 & 4 & 3 & 5/2 & 3 \\
Q$_{21}(3)$ & 835439488 & -1 & 3/2 & 7/2 & 4 & 3 & 7/2 & 4 \\
& 835439517 & -3 & 3/2 & 7/2 & 4 & 3 & 7/2 & 3 \\

R$_{2}(4)$ & 835459901 & -5 & 3/2 & 9/2 & 4 & 4 & 7/2 & 3 \\
\hline

R$_{21}(0)$ & 835433178 & 2 & 3/2 & 3/2 & 1 & 0 & 1/2 & 1 \\
& 835433245 & 2 & 3/2 & 3/2 & 2 & 0 & 1/2 & 1 \\
& 835433392 & 2 & 3/2 & 3/2 & 1 & 0 & 1/2 & 0 \\
\hline
 \hline
 \end{longtable}

\bibliography{library}

\end{document}